\documentclass[10pt,twocolumn,letterpaper]{article}

\usepackage{iccv}
\usepackage{times}
\usepackage{epsfig}
\usepackage{graphicx}
\usepackage{amsmath}
\usepackage{amssymb}
\usepackage{enumitem}
\usepackage{float}
\usepackage{caption}

\usepackage[pagebackref=true,breaklinks=true,letterpaper=true,colorlinks,bookmarks=false]{hyperref}

\iccvfinalcopy 

\def\iccvPaperID{9825} 

\ificcvfinal\fi
\begin{document}

\title{Explainable Artificial Intelligence in Retinal Imaging \protect\\ for the detection of Systemic Diseases}

\author{Ayushi Raj Bhatt \\
\and
Rajkumar Vaghashiya
\and
Meghna Kulkarni
\and
Dr Prakash Kamaraj
}

\maketitle

\begin{abstract}
Explainable Artificial Intelligence (AI) in the form of an interpretable and semiautomatic approach to stage grading ocular pathologies such as Diabetic retinopathy, Hypertensive retinopathy, and other retinopathies on the backdrop of major systemic diseases. The experimental study aims to evaluate an explainable staged grading process without using deep Convolutional Neural Networks (CNNs) directly. Many current CNN-based deep neural networks used for diagnosing retinal disorders might have appreciable performance but fail to pinpoint the basis driving their decisions. To improve these decisions' transparency, we have proposed a clinician-in-the-loop assisted intelligent workflow that performs a retinal vascular assessment on the fundus images to derive quantifiable and descriptive parameters. The retinal vessel parameters meta-data serve as hyper-parameters for better interpretation and explainability of decisions. The semiautomatic methodology aims to have a federated approach to AI in healthcare applications with more inputs and interpretations from clinicians. The baseline process involved in the machine learning pipeline through image processing techniques for optic disc detection, vessel segmentation, and arteriole/venule identification.
\end{abstract}

\section{Introduction}

Retinal health is essential in monitoring of individual health since it is a direct extension of the vasculature of the central nervous system. Changes in the retinal vasculature, such as thickening/widening of venules and narrowing of arteries, as well as tortuosity of the vessels and branching pattern of the daughter vessels, have proven to be significant biomarkers for the detection of pathologies that serve as a backdrop to systemic disorders \cite{42,31,34,27}.

\subsection{Impact of Deep Learning (DL) based models}

The most common pathology observed in retinal imaging is Diabetic retinopathy (DR), being one of the main causes of blindness at the global level, especially in the
working-age population \cite{7}. Experiencing severe grades of DR in the working age group can cause severe loss in productivity. Deep Learning has become a key tool in terms of
AI-assisted diagnosis of DR \cite{4,17}. There are also instances of Deep learning modules, such as CNNs that play a role in the classification of hypertensive retinopathy (HTR) \cite{43}. Features such as the narrowing of vessels in the retina, vascular
bleeding, and cotton wool spots play a role in the detection of retinopathy diseases of varied etiology. Deep learning has also helped improve the staging of diabetic retinopathy
conditions \cite{39}.

\subsection{Automated retinal AI for detection of diseases}

The diagnosis of the majority of retinopathies from the fundus images by clinicians is based on visual clues such as abnormal retinal vasculature morphology, bleeding, exudates, hemorrhages, and so on. Thus, computer vision greatly underpins the automatic detection of retinal conditions via image processing techniques such as blood vessel segmentation, classification \cite{3}, optic disc detection, and exudate localization. When combined, these intelligent algorithms can help detect ocular-cum-systemic conditions such as hypertension, diabetic retinopathy, and atherosclerosis. There have been extensive CNN-based researches such as on the automatic grading of retinal vessels \cite{29} and detection of lesions, microaneurysms, cotton wool spots, exudates, and retinal hemorrhages \cite{2,15,20}. These techniques have attained appreciable performance comparable to that of trained ophthalmologists. Further, various systems have been developed for specific disease detection and management, such as for hypertensive retinopathy \cite{1,23}.

\subsection{The explainable AI}

Explainable AI (XAI) are a modern-day example of having more accountable artificial intelligence systems where the results are open to interpretation, paving for better understanding by humans. XAI has been a conscious exploration in the realm of retinal imaging analysis as well \cite{24}. However, there is always uncertainty when it comes to deep learning-based models of retinal diseases \cite{38}. There is a constant need for an analysis system better understood by an ophthalmologist to avoid erroneous diagnoses. The difficulty is also experienced in advanced imaging systems such as retinal OCT diagnosis \cite{6}. There is a need to explore more on the domain with attributable risk factors with more qualitative and quantitative analysis \cite{37}.

\subsection{Diseases in focus}

Analysis of retinal vasculature could be extrapolated not just in ophthalmic conditions but in systemic diseases as well. The major systemic diseases that draw inference from the retinal vascular analysis are retinopathy, associated with the following conditions.

\begin{itemize}[label=, itemsep=-4pt]
    \item Diabetes \cite{13}  
\item Diabetic Nephropathy \cite{19} 
\item Chronic kidney disease \cite{44} 
\item Hypertension \cite{9} 
\item Cardiovascular diseases \cite{14} 
\item Neurocognitive diseases \cite{10} 
\item Infectious diseases due to immunodeficiency \cite{40,41} 
\item Inflammatory diseases \cite{16} 
\end{itemize}

\subsection{Retinal vascular parameters and retinopathies}

Many of the above-mentioned research utilize black-box models, i.e., the specific decisions taken by the model to conclude are not explainable. Thus, such systems resist adoption by clinicians despite their great performance.

Apart from image processing techniques and deep learning methods, there is a need to explore more meaningful parameters that contribute to specific conditions.
There have been significant findings that relate changes in the retinal vasculature to certain systemic diseases and retinopathies. Specific findings were noticed towards prediabetic and hypertensive conditions \cite{42}. Measurement of arterio-venous ratio has been especially useful in the detection of hypertensive conditions \cite{31,34}. Alterations in the retinal vascular caliber (RVC) are observed during routine retina examination during pre-pathological and pathological states \cite{19}. RVC has also indicated relevance in systemic disease conditions such as diabetic nephropathy and diabetic neuropathy.

Retinal vasculature parameters such as vascular tortuosity also have an impact on cardiovascular diseases. Retinal vascular tortuosity is associated with higher levels of blood pressure and HDL levels \cite{12}. Retinal image-specific signs such as arterio-venous nicking have linear correlations with cardiovascular disorders, and hypertension \cite{36}. Automated methods have been developed to quantify such parameters from color fundus images \cite{30,45}.

Fractal dimension (FD) is an indicative parameter that analyzes retinal vasculature as a fractal; the geometric parameterization helps to analyze the branching pattern of retinal vessels \cite{27}. Fractal dimension has been associated with significance in conditions like stroke \cite{11,26}. The vascular parameters-based approach would serve as an effective modus operandi for explainable AI in healthcare.

\section{Method}

\subsection{Diseases and corresponding datasets used in the study}

This study is focused on the detection and stage-grading
of three retinopathies, namely, Diabetic Retinopathy (DR),
Hypertensive Retinopathy (HTR) and Diabetic Macular
Edema (ME). Quite a few retinal datasets are available for
research purpose and additional analysis. Publicly available
datasets are especially important to analyze further, given
the additional confounding factors of varying datasets by
ethnicity, age, sex, and other etiological factors. Modern
imaging analysis methods through powerful computation
methods such as machine learning help in drawing additional
inferences \cite{22}.

The dataset used in the study for DR and Macular Edema:

\begin{itemize}[label=, itemsep=-4pt]
    \item IDRID \cite{33}
\end{itemize}

The datasets used in the study for HTR:

\begin{itemize}[label=, itemsep=-4pt]
    \item IDRID \cite{33}
\item AVRDB \cite{35}
\item JSIEC Images (Severe Hypertensive retinopathy category) \cite{21}
\end{itemize}

\section{Fundus Image Gradability Analysis}

\subsection{Grability analysis criteria}

Fundus photographs acquired from the respective
datasets were filtered based on the criteria of the visibility
of four major arterioles and venules. The images were
additionally analyzed to rule out the ones with prevailing
co-morbidities.

\subsection{Diabetic Retinopathy Dataset Analysis}

A total of 516 images filtered out from the IDRID dataset
were considered for DR analysis. The data analysis and
filtering process for DR dataset is shown in Fig. \ref{fig:1}.

The grade wise interpretation for multi-class classification
is based on the disease severity levels as specified:

\begin{figure}[H]
    \centering
    \includegraphics[width=0.5\textwidth]{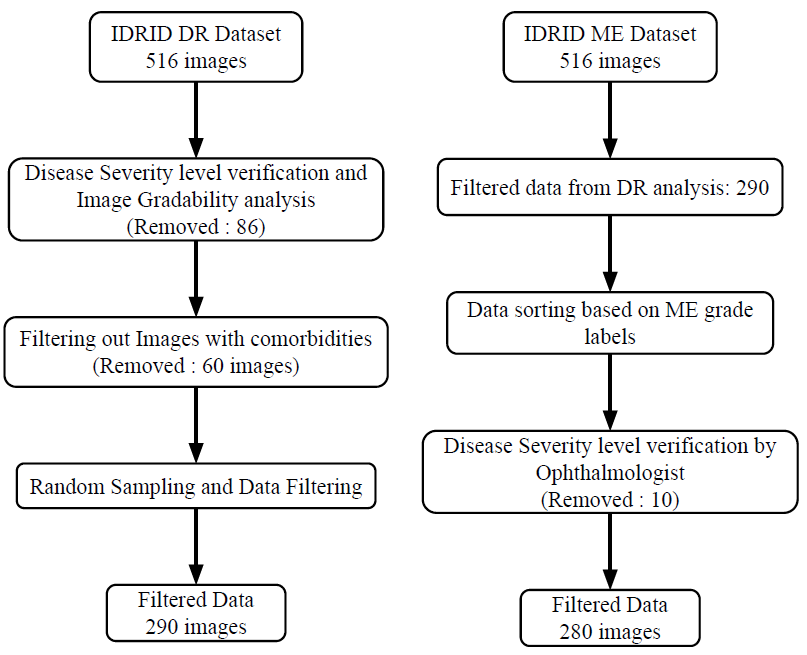}
    \caption{Filtering workflow for DR and ME datasets}
    \label{fig:1}
\end{figure}

\begin{table}[H]
    \centering
    \begin{tabular}{ccc}
    \hline
Severity Grade & Train Data & Test Data  \\ \hline
Grade 0 & 52 & 33 \\
Grade 1 & 20 & 5 \\
Grade 2 & 51 & 32 \\
Grade 3 & 40 & 19 \\
Grade 4 & 32 & 6 \\ \hline
    \end{tabular}
    \caption{Sample division in Diabetic Retinopathy Dataset}
    \label{tab:1}
\end{table}

\begin{itemize}[itemsep=-4pt, label=]
    \item Grade 0: Absence of DR
\item Grade 1: Mild DR
\item Grade 2: Moderate DR
\item Grade 3: Severe DR
\item Grade 4: Signs of proliferative DR
\end{itemize}

The final data sample distribution for DR is specified in
Table \ref{tab:1}. The binary grade classification was considered on
the presence or absence of the disease i.e. the Grade 0 signifies
the absence and the remaining higher grades signify
the presence of the disease.

\subsection{Diabetic Macular Edema Dataset Analysis}

516 images from the IDRID dataset for Diabetic Macular
Edema (ME) were filtered out and randomly sampled for
the image analysis. The data analysis and filtering process
for ME dataset is shown in Fig. \ref{fig:1}, and the final data sample
distribution for ME is specified in Table \ref{tab:2}. The grade
wise interpretation is based on the disease severity levels as
follows:

\begin{itemize}[itemsep=-4pt, label=]
    \item Grade 0: No visible exudates
\item Grade 1: Shortest distance between macula and
exudates $>$ one optic disc diameter
\item Grade 2: Shortest distance between macula and
exudates $\leq$ one optic disc diameter
\end{itemize}

\begin{table}[H]
    \centering
    \begin{tabular}{ccc}
    \hline
Severity Grade  & Train Data & Test Data \\ \hline
Grade 0 & 82 & 41 \\
Grade 1 & 14 & 9 \\
Grade 2  & 94 & 40 \\ \hline
    \end{tabular}
    \caption{Sample division in Macular Edema Dataset}
    \label{tab:2}
\end{table}

\subsection{Hypertensive Retinopathy Dataset Analysis}

\begin{figure}[H]
    \centering
    \includegraphics[width=0.5\textwidth]{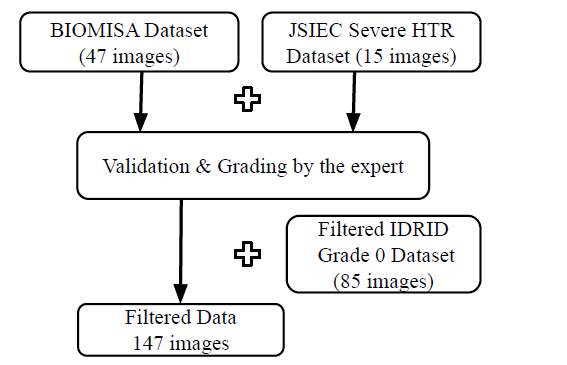}
    \caption{Filtering workflow for HTR dataset}
    \label{fig:2}
\end{figure}

A total of 66 hypertensive retinopathy images, combined
from the AVRDB and the JSIEC dataset, were used for
the analysis of higher grades of HTR. The control images
(Grade 0) were taken from the IDRID dataset after ensuring
the absence of comorbidities. The data collection, filtering
and selection procedure is specified as follows is specified
in Fig. \ref{fig:2}. The final data sample distribution for HTR is specified
in Table \ref{tab:2}.

Due to the lack of test data, 80-20 ratio split on the filtered
data was used for training and testing the model respectively.
The grade wise interpretation is based on the
following disease severity levels:

\begin{itemize}[itemsep=-4pt, label=]
    \item Grade 0: No visible abnormalities
\item Grade 1: Diffuse arteriolar narrowing
\item Grade 2: Grade 1 with focal arteriolar constriction
\item Grade 3: Grade 2 with retinal hemorrhage
\item Grade 4: Grade 3 with hard exudates, retinal
edema, and optic disc swelling
\end{itemize}

\begin{table}[H]
    \centering
    \begin{tabular}{cc}
    \hline
Severity Grade & Train Data \\ \hline
Grade 0 & 85 \\ 
Grade 1 & 11 \\
Grade 2 & 15 \\
Grade 3 & 20 \\
Grade 4 & 16 \\ \hline
    \end{tabular}
    \caption{Sample division in Hypertensive Retinopathy Dataset}
    \label{tab:3}
\end{table}

\section{Calculation of Retinal Vascular Parameters}

\begin{figure}[H]
    \centering
    \includegraphics[width=0.45\textwidth]{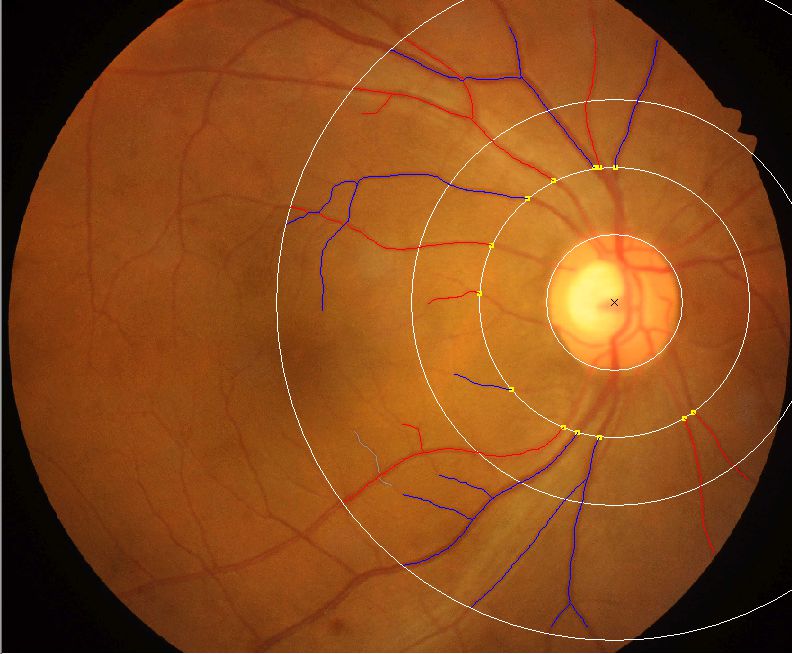}
    \caption{Vessel classification with retinal zones in SIVA}
    \label{fig:3}
\end{figure}

The filtered data was graded and retinal vessel characteristics
were extracted using the Singapore “I” Vessel Assessment
(SIVA) software \cite{25}. The semi-automated retinal
vascular network evaluation was based on the analysis
of vessels from the center of the optic disc and then to three
successive zones corresponding to 0.5 (zone A), 1 (zone B)
and 2 (zone C) optic disc diameters, as shown in Fig. \ref{fig:3}. The
parameters analyzed, per vessels, arterioles, and venules,
were as follows:

\begin{itemize}[label=, itemsep=-4pt]
    \item Central retinal arteriole equivalent (CRAE)
\item Central retinal venule equivalent (CRVE)
\item Arteriovenous ratio (AVR)
\item Fractal dimension (FD)
\item Mean vessel width (MW)
\item Standard deviation of vessel width (STDW)
\item Tortuosity (TORT)
\item Length to diameter ratio (LDR)
\item Branching coefficient (BC)
\item Asymmetry Factor (AF)
\item Branching angle (BA)
\item Asymmetry angle (AA)
\item Junction exponent (JE)
\item Number of branches \& number of first branches
\item Number of arterioles and venules
\end{itemize}

\begin{figure}[H]
    \centering
    \includegraphics[width=0.45\textwidth]{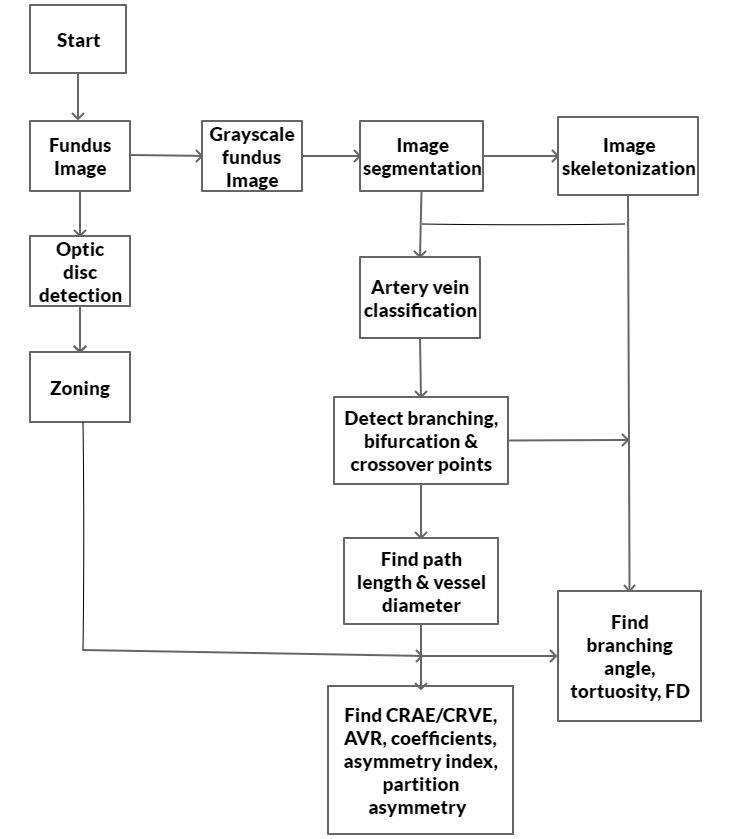}
    \caption{SIVA-based parametric quantification workflow}
    \label{fig:4}
\end{figure}

An $a$ beside the parameter, for \textit{e.g.} FD$a$, denotes arteriolar
measurement while a $v$, for \textit{e.g.} FD$v$ denotes venular
measurement. The derivation for the mentioned parameters
is specified in \cite{25}. The workflow for the parametric calculations
is specified in Fig. \ref{fig:4}.

\subsection{Stepwise Linear Regression Analysis}

A stepwise multi-linear regression analysis was performed
using IBM SPSS v23 to analyze the study parameters
and to check for multicollinearity in the input retinal
parameters.

\subsubsection{DR data analysis}

The input variables statistically significantly predicted the
DR severity stage, F(51, 238) = 3.118, p$<$.0005, R$^2$ =
.401. No variables were found that added statistically significantly
to the prediction, p$<$.05. As per the analysis, the
parameters Baa, AD2a, BAv, AAv, Bat, and AD2t were excluded
from the study. The details of the tests are provided
in the supplementary.

\subsubsection{HTR data analysis}

The input variables statistically significantly predicted the
HTR severity stage, F(51, 95) = 6.850, p$<$.0005, R$^2$ = .786.
The variables AVR (of both zone B and C), arteriolar MW
and BMW, and arteriolar NumLDR added statistically significantly
to the prediction, p$<$.05. As per the analysis, the
parameters BAa, AD2a, BAv, AAv, BAt, and AAt were excluded
from the study. The details of the tests are provided
in the supplementary.

\subsubsection{ME Data Analysis}

The input variables statistically significantly predicted
the ME severity stage, F(51, 228) = 1.567, p$<$.0005, R$^2$
= .260. The variables CRVE (of both zone B and C) added
statistically significantly to the prediction, p$<$.05. As per
the analysis, the parameters BAa, AD2a, BAv, AAv, AD1t,
and AD2t were excluded from the study. The details of the
tests are provided in the supplementary.

Most of the retinal parameters under consideration in this
study convey certain unique and causal information about
the retinal vasculature morphology in relation to the disease
being diagnosed. There is a possibility of spurious correlation
among the input parameters that might negatively interfere
and affect the multi-linear regression analysis.

\section{Machine Learning Study Design}

The experiment was conducted on a cloud-based compute
platform, Google Colab \cite{5}. Standard machine learning
algorithms for classification were evaluated on the data set,
mostly using scikit-learn \cite{32}, namely:

\begin{itemize}[label=, itemsep=-4pt]
    \item K-Neighbors Classifier (KNC)
\item Extreme Gradient Boosting (XGB) \cite{8}
\item Random Forest Classifier (RFC)
\item Multi-Layer Perceptron Classifier (MLP)
\item Decision Tree Classifier (DTC)
\item Gaussian Naive Bayes Classifier (GNB)
\item Gaussian Process Classifier (GPC)
\item AdaBoost Classifier (ABC)
\item Quadratic Discriminant Analysis (QDA)
\item Logistic Regression (LR)
\item Support Vector Classifier (SVC)
\end{itemize}

\begin{figure}[H]
    \centering
    \includegraphics[width=0.48\textwidth]{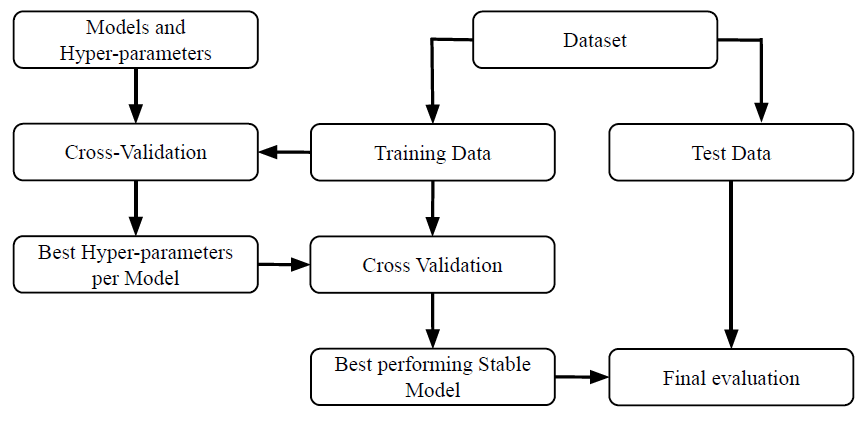}
    \caption{ML model evaluation workflow}
    \label{fig:5}
\end{figure}

For feature scaling, the data of each continuous parameter
is transformed using Min-Max scaling technique to normalize
the variance to a standard range i.e. between 0 and 1.
This is done to balance the effect of parameters, \textit{e.g.} CRVE
and CRAE, typically having higher values compared to the
rest, \textit{e.g.} AVR and FD. Following this, tSNE was utilized to view the class representations and check for inter-class
separability.

Cross-validation has been utilized to ensure consistent
performance with minimal tolerance in the accuracy. Nested
cross-validation on the training dataset was carried out using
4 inner folds for determination of the optimal model hyperparameters,
and 6 outer folds for determination of the model
performance and stability on the dataset. The top 4 performing
models were then evaluated on the test set to determine
the final model performance. The workflow of model hyperparameter
selection and evaluation is summarized in
Fig. \ref{fig:5}.

A weighted one-vs-rest Receiver Operating
Characteristics-Area Under the Curve (ROC-AUC) \cite{18}
 the scheme was used as the scoring criteria. ROCAUC
is a parameter-agnostic graphical performance metric for
binary classification problems used to depict how well a
discriminator model minimizes the overlap between the
two class distributions, the diseased and the non-diseased.
ROC is the probability curve derived from the evaluation of
the test samples while AUC denotes the area under that
curve. AUC$\leq$0.5 implies model performs no better than a
random guess, while AUC$>$0.5 implies the model is able to
discriminate well between the classes, with AUC of 1 being
the maximum possible score. It is adapted for multi-class
problems by pitting one class against the rest (one-vs-rest).

\vspace{6pt}

\textbf{Range of Hyper-parameters}

\vspace{6pt}

KNC: weights : [uniform, distance]

algorithm: [auto, ball tree, kd tree]

XGB: learning\_rate: [$10^{-3}$, $10^{-2}$, $10^{-1}$]

n\_estimators: [100, 120, 140]

RFC: max\_depth: [1, 2, 4, 8, 16, 32, 64, None]

n\_estimators: [8,9,10,11,12,13, 100, 120, 150, 200]

min\_samples\_split: [1,2,3]

MLP: hidden\_layer\_sizes: [10, 20, 40, 60, 80, 100, 120, 150, 200]

DTC: max\_depth: [1, 2, 4, 8, 16, 32, None]

min\_samples\_split: [1,2,3]

criterion: [gini, entropy]

GNB: var\_smoothing: [1e-9]

GPC: max\_iter\_predict: [25, 50, 75, 100, 120, 150]

ABC: n\_estimators: [10, 20, 40, 50, 70, 100, 120]

QDA: tol: [1e-4]

LR : max\_iter: [50, 100, 500, 100, 5000, 10000]

C: [$10^{-4}$, $10^{-3}$, $10^{-2}$, $10^{1}$, $10^{0}$, $10^{1}$, $10^{2}$]

SVC: C: [$10^{-4}$, $10^{-3}$, $10^{-2}$, $10^{1}$, $10^{0}$, $10^{1}$, $10^{2}$]

kernel: [poly, rbf, sigmoid]

\begin{figure}[H]
    \centering
    \includegraphics{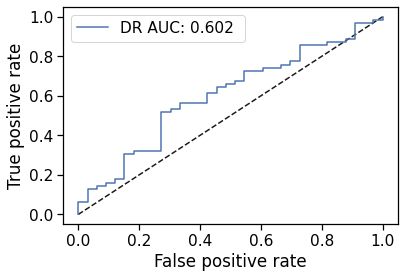}
\end{figure}

\begin{figure}[H]
    \centering
    \includegraphics{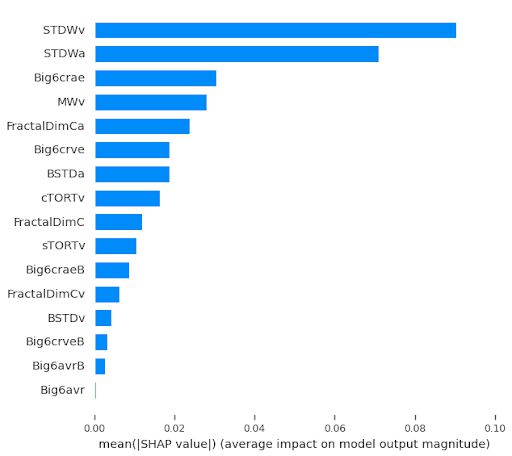}
    \caption{(a) DR Detection Results (b) DR Detection Feature Importance}
    \label{fig:6}
\end{figure}

\section{Classification Results}

\subsection{DR Detection Results}

ROC-AUC of 0.602 has been achieved for DR detection
as depicted in Fig. \ref{fig:6}. The best model is XGB with hyperparameters
learning\_rate: 0.01 and n\_estimators: 140.

For the explanation of the results, a game-theoretic approach
named SHAP \cite{28} is used to analyze the individual
impact of the input retinal features in the decision-making
process. The feature analysis for DR detection reveals that
the vessel width deviations (STDWa and STDWv) and the
central venular equivalent (CRVE) are the leading parameters
driving the decision on the detection of the disease.

\subsection{DR Stage Grading Results}

\begin{figure}[H]
    \centering
    \includegraphics{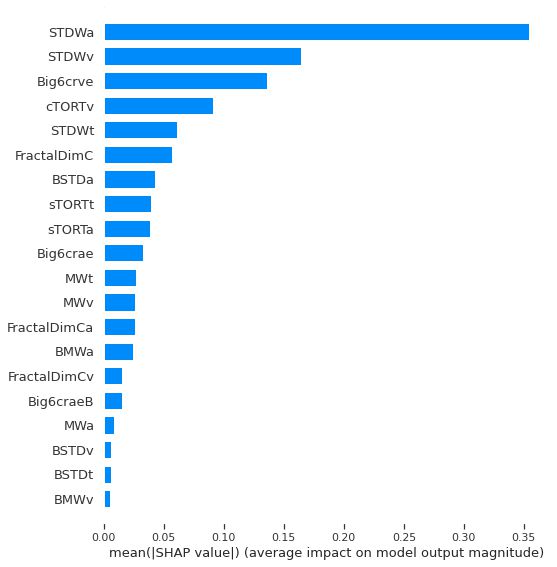}
    \caption{(a) DR Stage grading Results (b) DR Stage grading Feature
Importance}
    \label{fig:7}
\end{figure}

For DR stage grading, the performance of $\sim$0.67 has been
attained for stages 1 and 2, while ROCAUC of 0.812 and
0.863 has been attained for the most severe stages i.e. Grade
3 and 4 respectively as depicted in Fig. \ref{fig:7}. The best model
is Random Forest Classifier (RFC) with hyperparameters max\_depth: 8, min\_samples split: 3, and n\_estimators: 100.

\begin{figure}[H]
    \centering
    \includegraphics{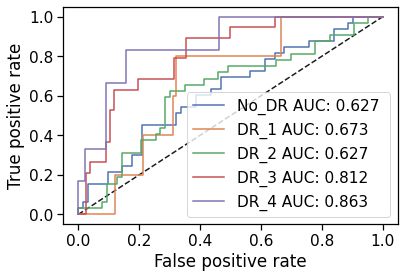}
\end{figure}

Consistent with the DR detection, the SHAP feature
analysis reveals that the vessel width deviations (STDWa
and STDWv) and the central venular equivalent (CRVE) are
the leading parameters driving the decision on the DR stagegrading
as well.

\subsection{HTR Detection Results}

\begin{figure}[H]
    \centering
    \includegraphics{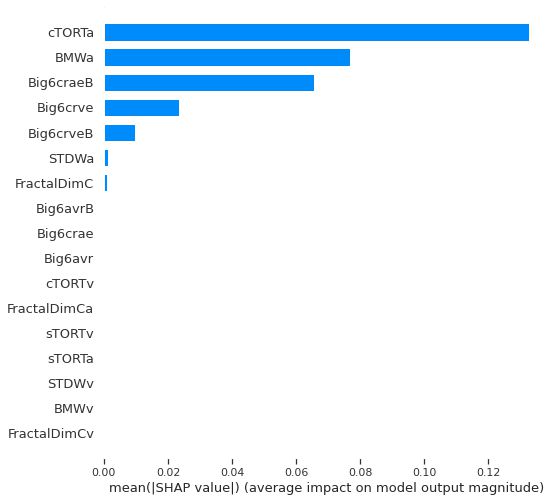}
    \caption{(a) HTR Detection Results (b) HTR Detection Feature
Importance}
    \label{fig:8}
\end{figure}

For HTR detection, high performance with ROCAUC
of 0.937 has been achieved as depicted in Fig. \ref{fig:8}. The best
model is XGB with hyperparameters learning\_rate: 0.01
and n\_estimators: 120. The SHAP analysis reveals that the
tortuosity (curvature tortuosity in this case) is the primary
deciding factor in HTR detection, and this fact has been
corroborated by past findings from retinal experts.

\begin{figure}[H]
    \centering
    \includegraphics{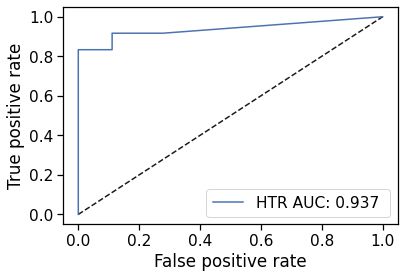}
\end{figure}

\subsection{HTR Stage Grading}

\begin{figure}[H]
    \centering
    \includegraphics{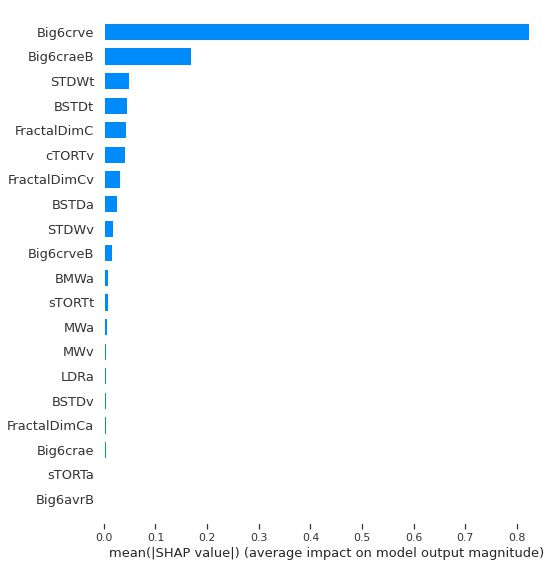}
    \caption{(a) HTR Stage grading Results (b) HTR Stage grading
Feature Importance}
    \label{fig:9}
\end{figure}

For HTR stage grading, a high-performance ROC-AUC
$\sim$ 0.95 has been obtained for stages 0,2 and 3, while
stage 1 has ROCAUC 0.877 and stage 4 has ROCAUC
0.721 respectively, as depicted in Fig. \ref{fig:9}. The best model
is Random Forest Classifier (RFC) with hyperparameters
max\_depth: 2, min\_samples\_split: 2, and n\_estimators: 100.
The SHAP feature analysis reveals that the central venular
and arteriolar equivalents as well as parameters like fractal
dimension jointly contribute towards the stage-grading of
the disease.

\begin{figure}[H]
    \centering
    \includegraphics{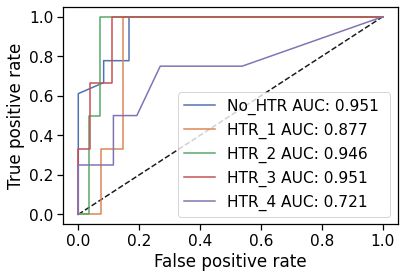}
\end{figure}

\subsection{ME Detection Results}

\begin{figure}[H]
    \centering
    \includegraphics{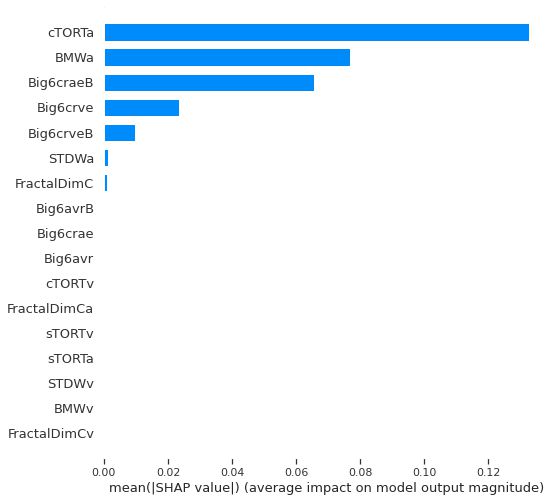}
    \caption{(a) ME Detection Results (b) ME Detection Feature
Importance}
    \label{fig:10}
\end{figure}

The performance in ME is similar to that achieved in
DR detection. For ME detection, ROC-AUC of 0.602
has been achieved for DR detection as depicted in Fig. \ref{fig:10}.
The best model is XGB with hyperparameters learning\_rate:
0.001 and n\_estimators: 140. The SHAP analysis reveals
that the detection depends on a lot of factors such as central vessel equivalents(Zone B CRAE and CRVE), arteriolar
vessel tortuosity (cTORT), and mean arteriolar width (Zone
B MWa).

\begin{figure}[H]
    \centering
    \includegraphics{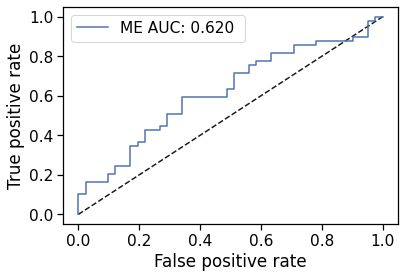}
\end{figure}

\subsection{ME Stage Grading}

\begin{figure}[H]
    \centering
    \includegraphics{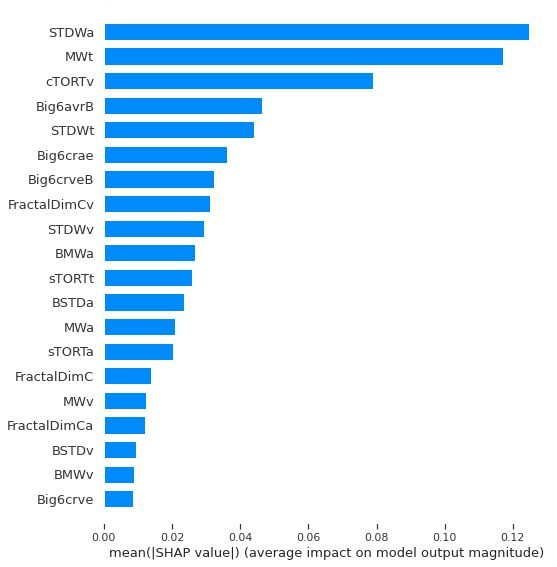}
    \caption{(a) ME Stage grading Results (b) ME Stage grading
Feature Importance}
    \label{fig:11}
\end{figure}

For ME stage-grading, the performance of ROC-AUC
$\sim$0.61 for Grades 0 and 1, while a ROC-AUC of 0.468
for Grade 2 is achieved as depicted in Fig. \ref{fig:11}. The best
model is AdaBoost Classifier (ABC) with hyperparameters
n\_estimators: 40.

Similar to the ME detection, the SHAP feature analysis
reveals that the ME stage grading jointly depends on factors
such as central vessel equivalents (CRAE and CRVE),
vessel tortuosity (cTORT), and fractal dimension (Zone C
venular FD) as observed above.
The final results of the study are shown in Table \ref{tab:4}. This
the study explores how the changes in vascular morphology,
when quantified and employed, could serve as the much-needed explanatory biomarkers that help in the understanding of the decision-making process as well result interpretation
of the intelligent models.

\begin{figure}[H]
    \centering
    \includegraphics{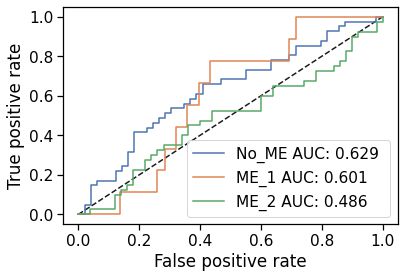}
\end{figure}

\begin{table}[H]
    \centering
    \begin{tabular}{ccccc}
    \hline
Task  & Best  & Train Perf.  & Test Perf. & Train Time \\
& Model & ROCAUC & ROCAUC & in min \\ \hline
\multicolumn{5}{c}{Disease Detection} \\ \hline
DR & XGB & .776$\pm$.095 & .602 & 8.077 \\
HTR & XGB & .986$\pm$.027 & .937 & .048 \\ 
ME & XGB  & .650$\pm$.115 & .620 & 0.986 \\ \hline
\multicolumn{5}{c}{Disease Grading} \\ \hline
DR & RFC & .799$\pm$.049 & .721 & 2.675 \\
HTR & RFC & .976$\pm$.009 & .889 & 1.877 \\
ME & ABC & .486$\pm$.077 & .572 & 2.735 \\ \hline
    \end{tabular}
    \caption{Experiment Results}
    \label{tab:4}
\end{table}

\section{Conclusion and Future work}

Despite having a relatively smaller dataset to work with,
a multivariate approach of specific retinal vascular parameters
have shown a reasonable association with retinal signs
caused due to systemic pathologies. Deep learning
methods need a larger dataset to work, needing at least
2000 images to have reasonable classifier results. The retinal
vascular parameters accounting for reliable parameters
showing specific associations to systemic diseases, we could
have better results on larger datasets with dedicated preprocessing
and filtering for other comorbidities that could
affect the co-variate results. Better literature review and
clinical studies for the specific parameters in tandem with
other secondary clinical parameters will lead to better results
and thus paving the way for explainable AI of retinal
vascular analysis.

{\small
\bibliographystyle{ieee}
\bibliography{egbib}
}

\end{document}